\documentclass[twocolumn]{svjour3}          % twocolumn

\smartqed  % flush right qed marks, e.g. at end of proof
\usepackage{amsmath}
\usepackage{eucal}
\usepackage{amsfonts}
\usepackage{graphicx}
\usepackage{multirow}

\journalname{arXiv}
\begin{document}

\title{Risk Measurement, Risk Entropy, and Autonomous Driving Risk Modeling} %\thanks{Grants or other notes
%about the article that should go on the front page should be placed here. General acknowledgments should be placed at the end of the article.}

\author{Jiamin Yu}        % \and Second Author %etc.

\institute{Jiamin Yu 
		\at Shanghai Lixin University of Accounting and Finance, \\
			Shangchun Road 995, Pudong, Shanghai 201209, China  \\
              \email{iambabyface@hotmail.com}           %  \\
}

\date{Received: date / Accepted: date}
\maketitle

\begin{abstract}
It has been for a long time to use big data of autonomous vehicles for perception, prediction, planning, and control of driving. Naturally, it is increasingly questioned why not using these big data for risk management and actuarial modeling. This article examines the emerging technical difficulties, new ideas, and methods of risk modeling under autonomous driving scenarios. \\
This paper first discusses the changes in risk modeling under autonomous driving scenarios. By information theory, a communication system framework of risk is proposed. Thus, the risk modeling problem is transformed into a problem of information transmission from real-world crash risk to estimated crash risk. Consequently, the optimization principle of modeling is to maximize the mutual information between real risk and estimated risk. In order to improve the risk identification resolution, a new-type risk time metrics, named TTA, is defined to satisfy that risk events are measurable. Furthermore, the expected value of TTA, defined as risk entropy formally, is mathematically proven to be Shannon's information entropy. It is the first time that actuarial science, information theory, and traffic engineering reach consensus on surrogate risk measurement. With the assumption that the driving process is a stochastic process of TTA, a microscopic states risk quantitative model is proposed. Consequently, the probability and time of a road accident are exactly the exit distribution and exit time of the Markov chain. Furthermore, the transition probability matrix of the risk model is extended to vary with real-time traffic flow and stimulus-response performance. Finally, driving data of car-following is used to illustrate the new model.\\
Compared with the traditional risk model, the novel model is more consistent with the real road traffic and driving safety performance. More importantly, it provides technical feasibility for realizing risk assessment and car insurance pricing under a computer simulation environment.
\end{abstract}

\section{Introduction}
\label{intro}
According to the definition of Wikipedia, an autonomous vehicle (AV), also known as a self-driving car, connected and autonomous vehicle (CAV), is a vehicle that is capable of sensing its environment and moving safely with little or no human input. In May, 2021, PR Newswire reported “The global autonomous car market is expected to reach 1383.89 billion dollar in 2025 at a CAGR of 14\%.” Since 2019, the newly launched car models of major automobile manufacturers have been AVs above Level 2 (see SAE J 3016-2018), and autonomous vehicles are becoming more and more popular.
\paragraph{}
Many proponents of self-driving, including users, automakers, technicians and specialists, claim that autonomous vehicles can eliminate most crashes and promote traveling safety greatly, hence insurance companies should offer special premium discount. However, in reality, most actuaries and insurers associate self-driving with huge risks and uncertainties. AVs are not only ineligible for premium discount, but also charged for extra premium. The pros and cons of premium discount result from disagreements and contradictions on premium pricing of AVs. 
\paragraph{}
Nowadays, the premium pricing conflicts turn into an outbreak phase. In August, 2019, due to dissatisfaction with premium discrimination, Tesla Motor launched Tesla Insurance, with higher rate discount and flexible payment modes. Although, Tesla considers Tesla autonomous cars safer than others, how to make pricing for Tesla's active safety and self-driving precisely, still a big technical challenge which Tesla must overcome. In July, 2020, Tesla CEO Elon Musk publicly called for “revolutionary actuaries” to join him to create a revolutionary Tesla Insurance. Thus it can be seen that, revolutionary actuarial risk modeling is the most significantly advanced technology for autonomous vehicle insurance.
\paragraph{}
Obviously, advanced actuarial models are essential for autonomous driving scenarios. However, there are barely few researches on actuarial modeling of autonomous vehicle. Most of the researches on the accident risk of AVs belong to the fields of vehicle engineering or traffic engineering. Several studies on autonomous car insurance mainly focus on the experience analysis of claims for the accident prevention effects of active safety functions, such as \cite{hldi2011Volvo}\cite{hldi2012Volvo}\cite{isaksson-hellman2015real}. 
\paragraph{}
For the gaps in risk modeling under autonomous driving scenarios, this article attempts to make an exploratory research on autonomous vehicle risk modeling approaches. The rest of this paper is organized as follows: Setion~\ref{sec:1} discusses the changes in data and requirements of actuarial models under self-driving scenarios; Setion~\ref{sec:2} introduces a new research framework: an autonomous driving risk (information) system. A new-type risk time metrics TTA is defined to satisfy that risk events are measurable, and information-theoretical optimization principle of risk modeling is proposed; Setion~\ref{sec:3} discretizes the state space of risk time distance TTA. Since the TTA process can be modeled as a discrete Markov chain according to the decision-making logic of autonomous driving, the probability and time of a road accident are exactly the exit distribution and exit time of the Markov chain; In Setion~\ref{sec:4}, car-following simulation experiments show how traffic flow and autonomous driving safety parameters affect rear-end collision risk quantitatively; The paper closes with its Conclusions section.
\section{Changes in Autonomous Driving Risk Modeling}
\label{sec:1}
\subsection{Big Data Upgrades from Autonomous Driving}
For a long time, both transportation industry and automobile insurance industry insist that crashes are the results of the combined effect of three types of risk elements: human, vehicle and road environment \cite{habibovic2013driver} \cite{yu2004using}. Automobile insurance actuaries have been making great efforts to collect more and more risk factor data to reflect the risk of accidents. The major risk factors, capturing potential crashes, include driver, vehicle and traffic environment related factors (as Table~\ref{tab:1}).
\begin{table}[h]
	\centering 
\caption{The major risk factors of human, vehicle and traffic environment.}
\label{tab:1}       % Give a unique label
\begin{tabular}{p{6em}p{17em}}
\hline\noalign{\smallskip}
Categories of risk factors & Risk factors\\
\noalign{\smallskip}\hline\noalign{\smallskip}
Human(driver) & Driving ability and behavior; Personal traits; Personal status (e.g. fatigue, distraction) \\
Vehicle & Equipment function failures; Maintenance status; Vehicle design defects  \\
Traffic-environment & Road geometry and design; Traffic conditions; Light and sight conditions; Weather conditions  \\
\noalign{\smallskip}\hline
\end{tabular}
\end{table}
\paragraph{}
With the popularization of automotive electronics technology, vehicle’s manipulation data (e.g. accelerator/ brake/ steering) and operating status data (e.g. mileage/ speed/ malfunction) can be easily collected through the vehicle bus. Nowadays, many insurance companies use mature telematics data collection to develop mileage-based car insurance, which is also called PAYD (Pay-As-You-Drive). Once telematics technology is upgraded to autonomous driving technology, high-dimensional, high-frequency, and spatial-temporal self-driving big data is available for risk modeling. AVs are usually equipped with high-definition(HD) cameras, radars and motion sensors, which can collect high-frequency, real-time video and various motion data (as Table~\ref{tab:2}). These multi-sensor data is processed by data analysis and data fusion algorithms to yield multiple object detection data, such as {\it traffic participants} (pedestrians/cyclists/motors), {\it driving zones, traffic signals and signs, road marking, obstacles and other objects}.
\begin{table}[h]
	\centering 
\caption{The sensor data and detection data of autonomous vehicles.}
\label{tab:2}       
\begin{tabular}{p{7em}p{6em}p{10em}}
\hline\noalign{\smallskip}
Sensors & Sensor data & Detection data\\
\noalign{\smallskip}\hline\noalign{\smallskip}
Motion sensors (GPS/IMU) & Position speed /Angular velocity & \multirow{3}{10em}{Identification outcomes of traffic participants, driving zones, traffic signals and signs, road marking, obstacles and other objects}\\
HD cameras & 2D photo /video \\
Radar (LiDAR/ millimeter-wave radar) & 3D distance vector \\
\noalign{\smallskip}\hline
\end{tabular}
\end{table}
By using these autonomous driving data, automobile and insurance companies can technically restore and reconstruct driving scenarios (including traffic conditions and driving actions), which provide ex-ante information beyond police accident reports(PARs) and site survey reports. For example, Tesla Services Center can retrieve historical data to aid insurance companies uncovering the accident process. Moreover, these autonomous driving data also makes it possible to create extra microscopic risk indicators for actuarial risk models.
\subsection{New Requirements and Problems for Autonomous Driving Risk Modeling}
Apart from matching big data, the autonomous driving actuarial risk model should also meet other novel concomitant requirements (challenges).
\paragraph{}
Firstly, the risk model should meet the insurance pricing requirements of more risk classifications. Pricing big data means the greater feasibility of numerous risk classifications. Classes of autonomous driving risk may be distinguished by various risk factors, such as speed, mileage, travel time (rush hour or slack hour), road grade (highway or street), etc. Numerous risk classifications are likely to cause {\it ecological paradoxes}. Traditional automobile insurance actuarial model is an up-bottom statistical inference approach. It usually aggregates the experience data at company level, and then applies the aggregate statistical regression effects of pricing factors to each categorical risk classification. For autonomous driving big data with numerous risk classifications, the traditional up-bottom statistical inference is prone to cause ecological fallacies. For example, Reference \cite{davis2004possible} shows that the {\it driving speed} is prone to cause ecological paradoxes in the risk of car-pedestrian crash, and the {\it vehicle speed} is very popular pricing factor of telematics car insurance. Therefore, how to ensure the pricing accuracy of risk classifications (i.e. eliminating ecological paradoxes) is the first problem to be solved by risk modeling.
\paragraph{}
Secondly, the risk model should have a “predictive” function to capture novel risks emerged from advancing autonomous driving technology. Autonomous driving technology is constantly developing and upgrading, and the self-driving function kit of AVs may be upgraded every year (in compliance with Moore's Law). Practically, the experience rating actuarial model requires several years' experience data, which will produce a time lag effect of risk matching. Assuming that the safety of autonomous driving is increasing over time, the actuarial risk of experience rating will always be greater than the current real-world risk. For autonomous vehicle, the risk reduction effect of autonomous driving, is the most typical actuarial problem. Therefore, how to predict premium discount on self-driving function is the second problem to be solved by risk modeling.
\paragraph{}
Finally, the risk model should have a “diagnostic” function to correctly identify the true “causes” of accident risks. Only when a risk model can identify the cause-and-effect of risks, it is eligible to work on disaster prevention and safety enhancement. Because effective countermeasures of disaster prevention and safety enhancement must be based on the correct identification of risk causality. The National Highway Safety Administration (NHTSA), as the highest authority for automobile safety, defined the cause of a traffic accident as: “{\it Defect, if there is no such defect, there will be no accident (deficiency without which the accident would not have occurred)}.” \cite{treat1979tri}. NHTSA’s definition implies a cause-and-effect inference, that is, if the cause of the accident is eliminated, no accident will occur. This causal inference is called counterfactual inference in Computer Science and Causal Inference. The opposite of accident cause, means accident cause had not happened yet, is “counterfactual”. On the contrary, the crash cause event actually happened and finally led to an accident. Thus it can be seen that, a counterfactual cause is very useful not only for finding true cause, but also for designing countermeasures (i.e. active safety and passive safety designs). Therefore, how to strictly identify the cause of the accident risk is the third problem to be solved by risk modeling.
\paragraph{}
According to Pearl's causal inference theory \cite{pearl2018theoretical}, the above “predictive” and “diagnostic” problems both are causal inference problems. Seriously, the “predictive” problem is an interventional causal inference problem, such as “What is my premium discount if subscribing self-driving functions?”; the “diagnostic” problem is a counterfactual causal inference problem, such as “What happens if jaywalking pedestrian had not been overlooked?”. Table~\ref{tab:3} shows the common inference problems of autonomous driving car insurance.
\begin{table}[h]
	\centering 
\caption{Two types of cause-effect questions to be answered by actuarial risk models.}
\label{tab:3}      
\begin{tabular}{p{5em}p{5em}p{13em}}
\hline\noalign{\smallskip}
Type & Concern & Question examples \\
\noalign{\smallskip}\hline\noalign{\smallskip}
Counter-factual & Effect & What if? /What if I upgrade the autopilot kit, will my premium drop? \\
Vehicle & Causes & Why? /What if the automatic emergency brake had not been disable?  \\
\noalign{\smallskip}\hline
\end{tabular}
\end{table}
Therefore, enhancing the interventional and counterfactual causal inference capabilities of risk models is an important goal of risk modeling.
\section{Methodology}
\label{sec:2}
\subsection{The Input-output System of Autonomous Driving Risk}
Nowadays, the most popular car insurance actuarial risk model is the experience rating model, which predicts future accidents (or losses) based on the historical accident (losses) experience of the insured vehicle. The experience rating model can be abstracted into the schematic diagram as below.
% For two-column wide figures use figure*
\begin{figure}[h]
\centering
  \includegraphics[width=0.45\textwidth]{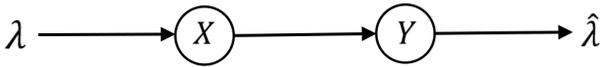}
\caption{The experience rating risk model (system).}
\label{fig:1}       % Give a unique label
\end{figure}
Where, $\lambda$ is the real world risk level(unknown hidden variable) and $\hat{\lambda}$ is the posterior estimated risk; $X$ and $Y$ are the underwriting risk variables (rate factors) and risk events (e.g. accident loss record) respectively. The core mathematical problem is to solve the conditional probability distribution of the risk level $\lambda$ under the observed conditions of $X$ and $Y$---$P(\lambda|X,Y)$.
\paragraph{}
The experience rating risk model implies a time-invariant assumption, that is, the future $\lambda$ and historical $\lambda$ remain unchanged. This significantly contradicts to our common sense that the risk of traffic accidents during rush hours is much higher than slack hours. It can be seen that traditional risk models cannot meet the requirements of precise pricing of autonomous driving risk, because even for the same self-driving car, the probabilities of accidents caused by different travel times and road sections (road environment) are completely different.
\paragraph{}
Apparently, for autonomous driving, accident risk is time-variant and road-variant. Assuming that the risk in Fig.~\ref{fig:1} is time-variant, the risk system transforms as follows
\begin{figure}[h]
\centering
  \includegraphics[width=0.45\textwidth]{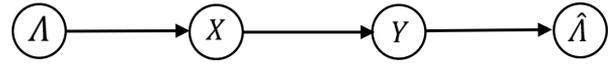}
\caption{The autonomous driving risk model (system).}
\label{fig:2}       % Give a unique label
\end{figure}
Where, the original $\lambda$ in Fig.~\ref{fig:1} was rewrite as $\Lambda$, which represents the time-varying autonomous driving risk level, and $\hat{\Lambda}$ is the corresponding estimated risk level. Mathematically, $\Lambda$ can be regarded as a function $\Lambda(t)$ of time $t$, or even a stochastic process $\Lambda_t$; $X$ and $Y$ are extended into risk input (risk factor variables) and risk output (risk events) from autonomous driving big data.
\paragraph{}
From the perspective of causal inference theory, the risk system in Fig.~\ref{fig:2} can be regarded as a delicate cause-effect mechanism, that is, changes in $X$ (e.g. vehicle malfunctions, road conditions) can directly lead to changes in risk event $Y$. Hence, $X$ and $Y$ can be regarded as {\it treatment variable} and {\it outcome variable} to solve abovementioned causal inference problems in Table~\ref{tab:3}.
\subsection{Risk Measurement}
\begin{figure}[h]
\centering
  \includegraphics[width=0.45\textwidth]{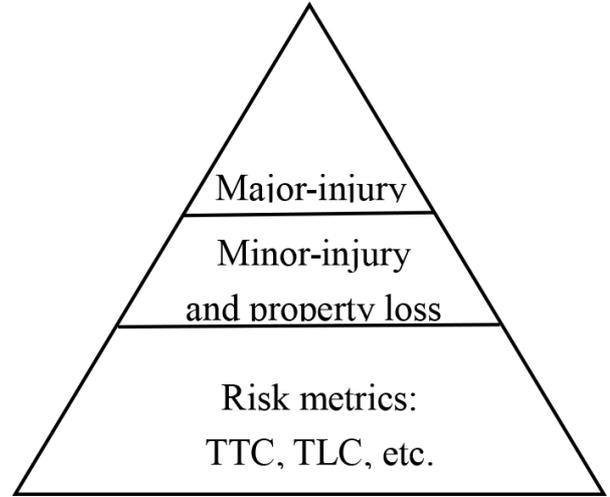}
\caption{The new Heinrich’s triangle of autonomous driving insurance.}
\label{fig:4}       % Give a unique label
\end{figure}

By theoretical definition, both {\it risk} (in risk theory) and {\it information} (in information theory) are uncertainties. Communication engineers and actuaries both estimate (or decode) signals with noise or errors (transmission code or accident risk), but their decoding accuracy (estimation) is completely different. Today, the decoding error probability of 4G and 5G is about $10^{-5}$, which is far beyond the reach of actuarial error probability. The reason lies in the high frequency signal measurement of wireless communication, usually up to millions of times per second, while the accident measurement of car insurance is usually several times per year. According to the Nyquist–Shannon sampling theorem in information theory \cite{moser2012a}, a large amount of high-frequency risk event information will be lost based on only the reported claim records, just like one can only see the tip of an iceberg above sea level, without knowing the hidden part below sea level. Therefore, the high-frequency measurement of risk through autonomous driving big data to discover the “hidden part” of car insurance risk, is a key engineering solution to solve the pricing accuracy problem under risk classifications.
\paragraph{}
In traffic engineering, surrogated safety measures, such as TTC \cite{hayward1972near}, TLC \cite{godthelp1984the}\cite{winsum2000a}, are often used to measure and quantify the collision risks. TTC and TLC are denoting time-to-collision and time-to-line crossing, quantifying front collision risks and side collision risks respectively. Take the most frequent rear-end collision risk between leading and following vehicles as an example, the TTC of the following vehicle equals to
\begin{equation}
TTC_f=\frac{X_l(t)-X_f(t)-l_f}{v_f(t)-v_l(t)}
\label{eq:1}
\end{equation}
Where, $X_l(t)$ and $v_l(t)$ denote the position and speed of leading vehicle, $X_f(t)$ and $v_f(t)$ denote the position and speed of following vehicle, $l_f$ is the length of following vehicle. Nowadays, those AVs with popular FCW (forward collision warning) and LWD (lane departure warning) functions, can calculate and output their real-time TTCs and TLCs. Hence, autonomous driving big data can feasibly provide lots of microscopic high-frequency risk measurements.
\paragraph{}
These new autonomous driving risk metrics will greatly expand the range of known risks for engineers and actuaries. In safety theory, Heinrich's Law, also known as the "300:29:1 theory", uncovers the relationship between the severity of accidents and the frequency of accidents \cite{heinrich1959industrial}. For automobile insurance, fatal or major-injury accidents, property loss and minor-injury accidents, unreported violations and unsafe behaviors also obey the Heinrich's law. The Heinrich’s Triangle of traditional automobile insurance accidents is shown as below.
\begin{figure}[h]
\centering
  \includegraphics[width=0.45\textwidth]{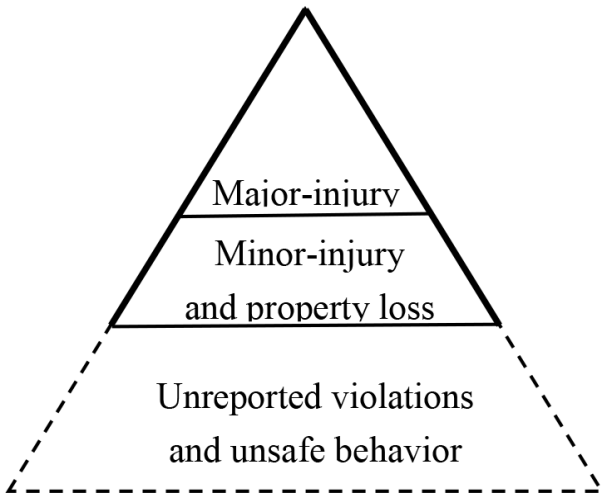}
\caption{The Heinrich’s triangle of traditional automobile insurance}
\label{fig:3}       % Give a unique label
\end{figure}
As shown in Fig.~\ref{fig:3}, the upper solid line triangle part denotes the known risks, and the lower dotted trapezoid part denotes the unknown risks. With the aid of autonomous driving risk metrics, engineers and actuaries can observe and measure more risks. The Heinrich’s Triangle of autonomous driving risks will become as Fig.~\ref{fig:4}.
Figure \ref{fig:4} shows that the past unknown risks, such as unreported violations and unsafe behaviors, will be knowable through autonomous driving risk metrics. In summary, autonomous driving big data will directly expand risk event $Y$ by risk measurement in engineering.
\subsection{Information-Theoretical Approach and Risk Entropy}
From the perspective of information theory, the expansion of risk event $Y$ as Fig.~\ref{fig:4}, significantly increase $Y$’s information (entropy). Furthermore, deem the risk system in Fig.~\ref{fig:2} as a communication information system, the autonomous driving actuarial problem can be transformed into a problem of information transmission (from real-world crash risk $\Lambda$ to estimated crash risk $\hat{\Lambda}$). Apply the mutual information theorem of information theory, the optimal risk modeling criterion is directly inferred to maximize the mutual information between $\Lambda$ and $\hat{\Lambda}$. Thus, the actuarial problem for precise risk pricing can be described as the following mathematical problem
\begin{equation}
\max\big(I(\Lambda,\hat{\Lambda})\big) \notag
\end{equation}
Where, $I(\Lambda,\hat{\Lambda})$ is the mutual information between $\Lambda$ and $\hat{\Lambda}$. With the data processing inequality, the following inequality is obtained
\begin{equation}
I(\Lambda,\hat{\Lambda}) \le I(Y,\hat{\Lambda}) \le \min(H(Y),H\bigl(\hat{\Lambda})\bigr)
\label{eq:2}
\end{equation}
Here, $H(Y)$ denotes the information entropy of $Y$. Thus, it can be obtained from (\ref{eq:2})  that, increasing $H(Y)$ is a prerequisite for maximizing $I(\Lambda,\hat{\Lambda})$. Obviously, autonomous driving big data in Fig.~\ref{fig:4} can be used to increase $H(Y)$. Such data processing (risk measurement) is called {\it refinement} in information theory. Therefore, risk metrics in Fig.~\ref{fig:4} actually play a role in risk refinement. That's because {\it only by refinement can mutual information be increased}.
\paragraph{}
Looking back at autonomous driving risk metrics, such as TTC and TLC, they are all time measurement to accident occurrence time. Naturally, they can be collectively called time-to-accident, abbreviated as TTA. TTA is the remaining time (negative value) before the accident. TTA is a random variable of risk state (or risk time distance), and different TTAs are mutually exclusive in probability. $P(TTA)$ denotes the probability of TTA. According to the Heinrich’s triangle in Fig.~\ref{fig:4}, the relationship between $TTA$ and $P(TTA)$ satisfies
\begin{equation}
P(TTA_a)>P(TTA_b)>0 \quad if \quad TTA_a < TTA_b < 0
\label{eq:3}
\end{equation}
 (\ref{eq:3}) shows $P(TTA)$ is a monotonic decreasing function of $TTA$.
\paragraph{}
Next, assume that the self-information of the risk event (its probability equals to $p_t=P(TTA=t)$ )is just the risk time distance $TTA$, that is
\begin{equation}
I(p_t)=t
\label{eq:4}
\end{equation}
Where, $I(p_t)$ denotes the self-information of the risk event, whose probability equals to $p_t$. According to (\ref{eq:3}) (i.e. risk time distance $TTA$) satisfying
\begin{enumerate}
	\item $I(p_t)$ is monotonically decreasing in $p_t$;
	\item $I(p_t)$ is a continuous function of  $p_t$ for  $0 < p_t \le 1$;
	\item {Only if $TTA$ obeys the exponential distribution exponential ($\lambda$), $p_t=f_{TTA}(t)=\lambda e^{\lambda t},\quad t \le 0,\lambda >0$ (in other words, accident count $N(t)$ is a Poisson process Poisson($\lambda$)), the Shannon's self-information of $p_t$ is
\begin{equation}
I(\lambda e^{\lambda t})=-\log_2^{\lambda e^{\lambda t}}=-\log_2^{\lambda}-\lambda t\log_2^e \notag
\end{equation}
Under this condition, risk time distance $TTA$ will be a linear transformation of Shannon's self-information.}
\end{enumerate}
Imitating Shannon’s expectation of self-information to obtain Shannon’s information entropy, expected risk time distance $TTA$ is defined as risk entropy. The risk entropy of $TTA$ is written as
\begin{equation}
H(TTA)=E[TTA]
\label{eq:5}
\end{equation}
As shown in (\ref{eq:5}), it is the first time that risk theory (actuarial science), information theory, and traffic engineering reach consensus on surrogate risk measurement $TTA$. Similar to Shannon’s information entropy, {\it risk entropy $H(TTA)$ represents the macro-state of risk}, that is, a macroscopic risk quantitative indicator.
\section{Modeling}
\label{sec:3}
\subsection{The Microstate of Risk Time Distance $TTA$}
In the previous section, $H(TTA)$ describes the macro-state of the risk process. In order to comply with the time-varying risk assumption, the autonomous driving risk system micro-state should be analysed to capture the mechanism of microscopic evolution during risk process. Hence, risk time distance $TTA$ is supposed to a stochastic process $\{TTA(t),0 \le t \le T\}$, and $TTA(t)$ is the state of the process at travel time $t$ ($T$ is the total travel time). Obviously, $\{TTA(t)\}$ (or abbreviated as $\{TTA_t\}$)is a continuous time stochastic process physically. Please note that there are two specific states, $TTA_0=-\infty$ and $TTA_t=0$, which represent parking state and crash state ($t$ is the time of the accident) respectively. For an accident-free trip, the status of $TTA$ at the beginning and end are: $TTA_0=TTA_T=-\infty$.
\paragraph{}
Due to the limitation of the data processing frequency of vehicle sensors and processors, the actually measured $TTA$ is a discrete time stochastic process, $\{TTA_n,0 \le n \le T/\delta,n\in \mathbb{N}\}$. Generally, $\delta$ is the sampling period of the autonomous driving data frame, for instance, $1/30$ seconds.
\subsection{State Space Discretization}
\begin{figure}[h]
\centering
  \includegraphics[width=0.45\textwidth]{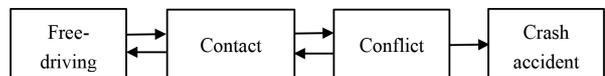}
\caption{Driving state switching schematic.}
\label{fig:5}       
\end{figure}

From the above analysis, it can be found that $TTA_t$ is a continuous variable, and its time interval is $(-\infty,0]$. In order to facilitate modeling and calculation, the continuous state space of $TTA_t$ needs to be discretized into certain discrete state space. Related research conclusions of driving state switching in traffic engineering, as shown in Fig.~\ref{fig:5}, will help discretize the state space.
\paragraph{}
In Fig.~\ref{fig:5}, the {\it free-driving} state represents no traffic participants nearby, or freely traveling without others’ influences, so free-driving state is the safest state; The {\it contact} state represents some traffic participants ahead contacted, and no accident will happen if keeping current driving style (e.g. car-following), so contact state is the second safest state; The {\it conflict} state represents collision-prone situation, in which two traffic participants approach each other and will crash if no evasive maneuver action is taken \cite{dijkstra2010do}\cite{dingus2006the}, and conflict severity is the best known safety indicator in traffic safety studies; The last is {\it crash accident} state. Please note that crash accident state is an {\it exit state}, that is, different from all other states, it cannot return to a safer conflict state.
\paragraph{}
For the $TTA$ safety measures of autonomous driving, such as TTC, TLC, etc., technical specifications and standards have designed corresponding time threshold values. $Thrd_{detect}$, $Thrd_{conflict}$, and $Thrd_{deadline}$ are usually common thresholds. $Thrd_{detect}$ indicates the minimum time headway (sensing range) that the vehicle sensor can detect at a high speed(e.g. $120km⁄h$). In theory, $Thrd_{detect}$ is enough to reserve time for accident avoidance operations. $Thrd_{conflict}$ is the critical time distance between safety and danger. Once $TTA$ less than $Thrd_{conflict}$, the vehicle computer will recognize traffic conflicts and execute evasive actions, such as braking, lane changing, etc. $Thrd_{deadline}$ is the minimum safe conflict time distance, which is the last moment to avoid accidents. Once $TTA$ less than $Thrd_{deadline}$, accidents are inevitable. Naturally, the continuous state space $(-\infty,0]$, can be segmented into three intervals,that is, $(-\infty,Thrd_{detect}]$, $(-Thrd_{detect},-Thrd_{deadline}]$, and $(-Thrd_{deadline},0]$, corresponding to the safest state, relatively safe and dangerous state, and the most dangerous accident state.
\paragraph{}
In order to increase the risk entropy in (\ref{eq:5}), it is necessary to further refine the above state space intervals. The refinement on $(-Thrd_{detect},-Thrd_{deadline}]$ is the most effective way to increase $H(TTA)$. Divide $(-Thrd_{detect},-Thrd_{deadline}]$ into $D$ intervals on average, and the span of each interval is equal to $\sigma=(Thrd_{detect}-Thrd_{deadline})/D,\sigma > \delta$(for instance, $\sigma=0.1s$ and $\delta=0.02s$). Thus, the discrete state space $\mathcal{S}$ becomes as 
\begin{multline}
\mathcal{S}=\{ (-\infty,Thrd_{detect}],(-Thrd_{detect},-Thrd_{detect}+\sigma],\\
		(-Thrd_{detect}+\sigma,-Thrd_{detect}+2\sigma],\dots,\\
		(-Thrd_{detect}+(D-1)\sigma,-Thrd_{deadline}],(-Thrd_{deadline},0]\}
\label{eq:6}
\end{multline}
As (\ref{eq:6}) shown, $\mathcal{S}$ has $D+2$ state elements (intervals). $\mathcal{S}_0$ and $\mathcal{S}_{D+1}$ are $(-\infty,Thrd_{detect}]$ and $(-Thrd_{deadline},0]$ respectively, also called state $0$ and state $D+1$. Similarly, state $i$, namely $\mathcal{S}_i$, is the interval $(-Thrd_{detect}+(i-1)\sigma,-Thrd_{detect}+i\sigma]$. Finally, the discrete state space $\mathcal{S}$ is simplified to
\begin{equation}
\mathcal{S}=\{ \mathcal{S}_0,\mathcal{S}_1,\dots,\mathcal{S}_D,\mathcal{S}_{D+1}\}
\label{eq:7}
\end{equation}
\subsection{Risk Transition Mechanism}
First review the driving decision-making logic of autonomous driving, which is directly related to {\it speed} ($u$), {\it traffic density} ($k$), and {\it flow} ($q$) in traffic dynamics. The relationship between the smooth speed $u_{smooth}$ and the traffic density $k$ under equilibrium conditions satisfies
\begin{equation}
u_{smooth}=u_e(k)
\label{eq:8}
\end{equation}
Where, $u_e(\cdot)$ is the equilibrium speed-density relationship function, and traffic density $k$ obviously depends on the real-time road conditions and speed limit $u_{max}$ \cite{nelson1995on}. Since self-driving cars usually adopt defensive driving ways (abandoning offensive driving ways), their driving decision-making logic is quite concise and clear, which can be directly described as the following algorithm flow chart.
\begin{figure}[h]
\centering
  \includegraphics[width=0.45\textwidth]{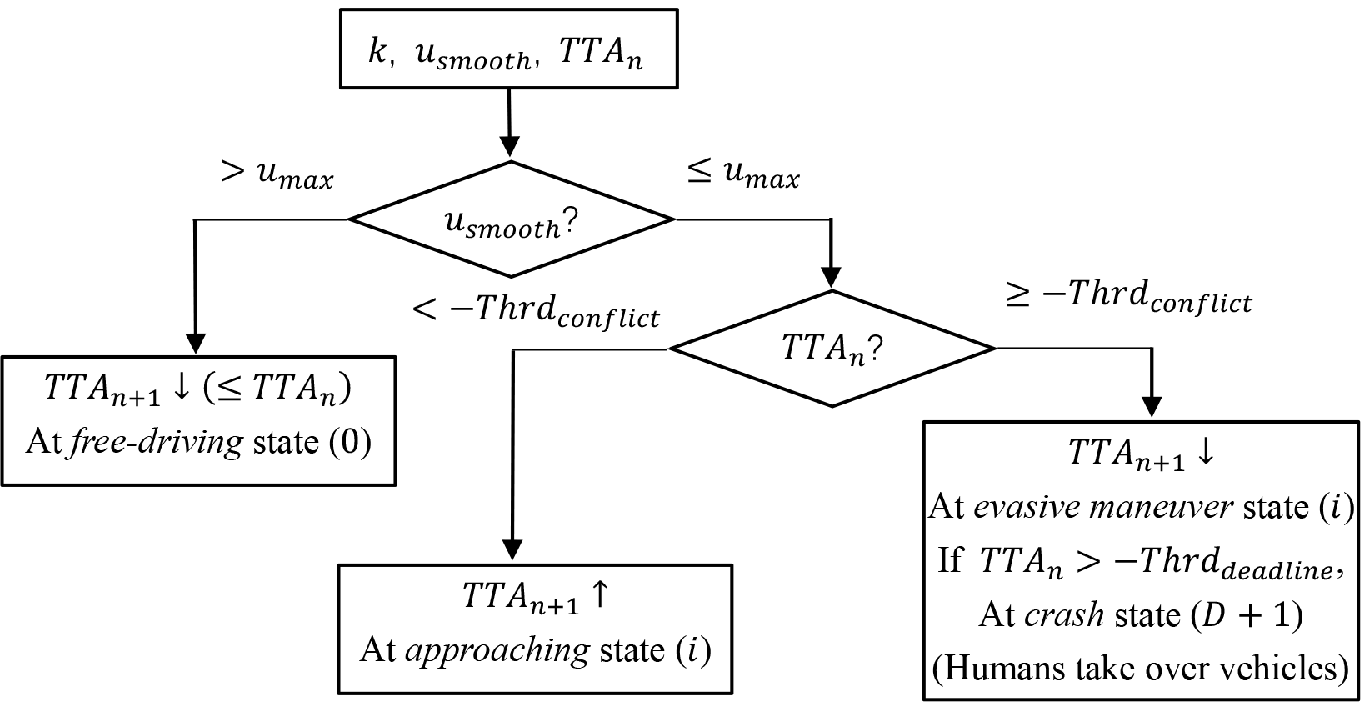}
\caption{An algorithm flow chart of autonomous driving decision-making logic.}
\label{fig:6}       
\end{figure}
As shown in Fig.~\ref{fig:6}, each state of discrete state space $\mathcal{S}$ in (\ref{eq:7}), may be covered by if-else statements. In particular, the approaching state ($i$) and evasive maneuver state ($i$) in Fig.~\ref{fig:6}, are the risk tension and relaxation state respectively.
\paragraph{}
On the basis of the flow chart in Fig.~\ref{fig:6}, it can be deduced that $\{TTA_n\}$ is a discrete Markov chain. And the transition probability matrix is obtained from the decision logic in Fig.~\ref{fig:6}.
\begin{equation}
\mathbf{D}=
\begin{bmatrix}
p_0   & q_0   & \quad  & \quad & \quad & \quad & \quad & \quad & \quad \\
0     &   0   &     1  & \quad & \quad & \quad & \quad & \quad & \quad \\
\quad & \ddots & \ddots & \ddots & \quad & \quad & \quad & \quad & \quad \\
\quad & \quad  & 0	  & 0      & 1     & \quad & \quad & \quad & \quad \\
\quad & \quad & \quad  & 0	     & 1      & 0     & \quad & \quad & \quad \\
\quad & \quad & \quad  & \quad & 1      & 0     & 0     & \quad & \quad \\
\quad & \quad & \quad & \quad & \quad  & \ddots & \ddots & \ddots & \quad \\
\quad & \quad & \quad & \quad & \quad & \quad   & 1      &  0     & 0  \\
\quad & \quad & \quad & \quad & \quad & \quad   & \quad  &  0     & 1  \\
\end{bmatrix}
\begin{array}{c}
0 \\ 1 \\ \vdots \\ c-1 \\ c \\ c+1 \\ \vdots \\ D \\ D+1
\end{array}
\label{eq:9}
\end{equation}
Where, $\mathbf{D}$ is a $D+2$ dimensional sparse matrix, and all its non-zero elements are shown in (\ref{eq:9}). Rows $1$ to $c-1$ of $\mathbf{D}$ show the tension ({\it approaching}) process of state ($i$), and rows $c+1$ to $D$ of $\mathbf{D}$ show the relaxation ({\it evasive maneuver}) process of state ($i$). $p_0$ and $q_0$ equal the probabilities of $P(u_{smooth}>u_{max})$ and $P(u_{smooth} \le {u_max})$.
\paragraph{}
The transition probability matrix $\mathbf{D}$ is idealized. It assumes that there are no errors and delays in the decision-making process. In order to simulate a realistic autonomous driving system, an error probability $\alpha$ and a delay probability $\beta$ are introduced to the state tension and relaxation process, with supposing that $\alpha$ and $\beta$ remain unchanged, and $\gamma=\alpha+\beta$. Then, the modified risk state transition matrix is
\begin{equation}
\mathbf{D}=
\begin{bmatrix}
p_0(t)& q_0(t)& \quad  & \quad & \quad & \quad & \quad & \quad & \quad \\
\alpha& \beta &1-\gamma & \quad & \quad & \quad & \quad & \quad & \quad \\
\quad & \ddots & \ddots & \ddots & \quad & \quad & \quad & \quad & \quad \\
\quad & \quad  &\alpha  & \beta  & 1-\gamma & \quad & \quad & \quad & \quad \\
\quad & \quad & \quad  & \frac{\gamma}{2}& 1-\gamma &\frac{\gamma}{2}& \quad & \quad & \quad \\
\quad & \quad & \quad  & \quad &1-\gamma & \beta & \alpha & \quad & \quad \\
\quad & \quad & \quad & \quad & \quad  & \ddots & \ddots & \ddots & \quad \\
\quad & \quad & \quad & \quad & \quad & \quad   & 1-\gamma  &  \beta & \alpha  \\
\quad & \quad & \quad & \quad & \quad & \quad   & \quad  &  0     & 1  \\
\end{bmatrix}
\label{eq:10}
\end{equation}
Note the transition probability of state $0$, $p_0(t)$ and $q_0(t)$, vary with the traffic density $k(t)$ at time $t$.
\paragraph{}
In short, the risk state transition matrix $\mathbf{D}$ in (\ref{eq:10}), changes with the road condition density $k(t)$, the automatic driving system error rate parameter $\alpha$, and the stimulus-response speed parameter $\beta$.
\subsection{Exit Distribution and Exit Time of an Accident}
If one journey of the vehicle is regarded as the risk exposure unit of risk pricing, then the vehicle will have two exit states: safe {\it terminal} state and {\it unsafe} accident state (i.e., state $D+1$). The moment of arriving at the terminal state is written as $T_{terminal}$, which means that the journey is safely finished. Similarly, $T_{accident}$ is the moment of arriving at the accident state, which means that the journey is unsafely interrupted ($T_{accident}<T_{terminal}$). Thus, a safe journey is from the previous terminal state to the next terminal state (terminal state $\rightarrow$ state $0$ $\rightarrow \ldots \rightarrow$ terminal state), and an accidental journey is from the previous terminal state to the accident state(terminal state $\rightarrow$ state $0$ $\rightarrow \ldots \rightarrow$ state $D+1$). Rewrite the terminal state as state $D+2$, then $\mathbf{D}$ in (\ref{eq:10}) will be extended to
\begin{equation}
\mathbf{ED}=
\begin{bmatrix}
p_1   & p_2   & \quad   & \quad & \quad & \quad & \quad & \quad & \quad & p_3\\
\alpha& \beta &1-\gamma & \quad & \quad & \quad & \quad & \quad & \quad & \quad\\
\quad & \ddots &\ddots & \ddots & \quad & \quad & \quad & \quad & \quad & \quad\\
\quad & \quad  &\alpha & \beta  & 1-\gamma&\quad & \quad & \quad & \quad & \quad \\
\quad & \quad & \quad  & \frac{\gamma}{2}& 1-\gamma &\frac{\gamma}{2}& \quad & \quad & \quad & \quad \\
\quad & \quad & \quad  & \quad &1-\gamma & \beta & \alpha & \quad & \quad & \quad  \\
\quad & \quad & \quad & \quad & \quad  & \ddots & \ddots & \ddots & \quad & \quad \\
\quad & \quad & \quad & \quad & \quad & \quad   &1-\gamma&\beta &\alpha & \quad \\
\quad & \quad & \quad & \quad & \quad & \quad   & \quad  &  0   & 1    & 0 \\
\quad & \quad & \quad & \quad & \quad & \quad   & \quad  &\quad & 0    & 1 \\
\end{bmatrix}
\label{eq:11}
\end{equation}
Where, $p_3$ equals to the probability $P(t=T_{terminal})$, $p_1$ equals to the probability $P(u_{smooth}>u_{max}|t<T_{terminal})$, and $p_2$ equals to the probability $P(u_{smooth}\le u_{max} |t<T_{terminal})$. It can be seen from the new transition probability matrix $\mathbf{ED}$ that, state $D+1$ and state $D+2$ are {\it recurrent states}, and state $0,1,\dots,D$ are {\it transient states}. Furthermore, state $D+1$ and state $D+2$ are two exit states, especially, the occurrence frequency of state $D+1$ (i.e., the inverse of {\it exit time}) directly determines the net risk premium.
\paragraph{}
First define the accident event as: $T_{accident}<T_{terminal}$ during a journey, and define the probability of exiting with an accident from each current state $x$ as $h(x)$. $h(x)$ equals to
\begin{equation}
h(x)=P_x(T_{accident}<T_{terminal})
\label{eq:12}
\end{equation}
Obviously, for state $D+2$, since $T_{accident}>T_{terminal}$, so $h(D+2)=0$; for state $D+1$, since $T_{accident} \ll T_{terminal}$, so $h(D+1)=1$. For other transient states $0,1,\dots,D$, use the exit distribution theorem \cite{durrett1999essentials} to get
\begin{equation}
h(x)=\sum_y{p(x,y)h(y)} \qquad x,y \in0,1,\dots,D+2
\label{eq:13}
\end{equation}
Where $p(x,y)$ is the element of state $x$ row and state $y$ column in matrix $\mathbf{ED}$.
\paragraph{}
The next {\it exit time} problem is also very similar. Define the time between the current state $x$ and the accident state $D+1$ as $g(x)$. Obviously, for state $D+1$, $g(D+1)=0$. For other states except state $D+1$, using the exit time theorem \cite{durrett1999essentials}, then have
\begin{equation}
g(x)=1+\sum_y{p(x,y)g(y)} \qquad x \in 0,1,\dots,D,D+2
\label{eq:14}
\end{equation}
(\ref{eq:14}) is equivalent to
\begin{equation}
g(x)=E_x(T_{accident})
\label{eq:15}
\end{equation}
Where, $E_x(T_{accident})$ represents the expected value of time from state $x$ to the accident state $D+1$ for the first time. 
\paragraph{}
It must be noted that $g(x)$ directly depends on the specifications of the automatic driving system: $\alpha$ and $\beta$, function settings: $c$, and real-time road conditions: $k(t)$ and $u_{max}$. Under the condition that the parameters of the automatic driving system remain unchanged, increasing the $TTA$ safety threshold of state $c$ or reducing the traffic density $k$, can prolong the arrival time of  accidents (i.e., reduce the probability of the accident).
\section{Simulation Experiment}
\label{sec:4}
\begin{figure*}[h]
\centering
  \includegraphics[width=\textwidth]{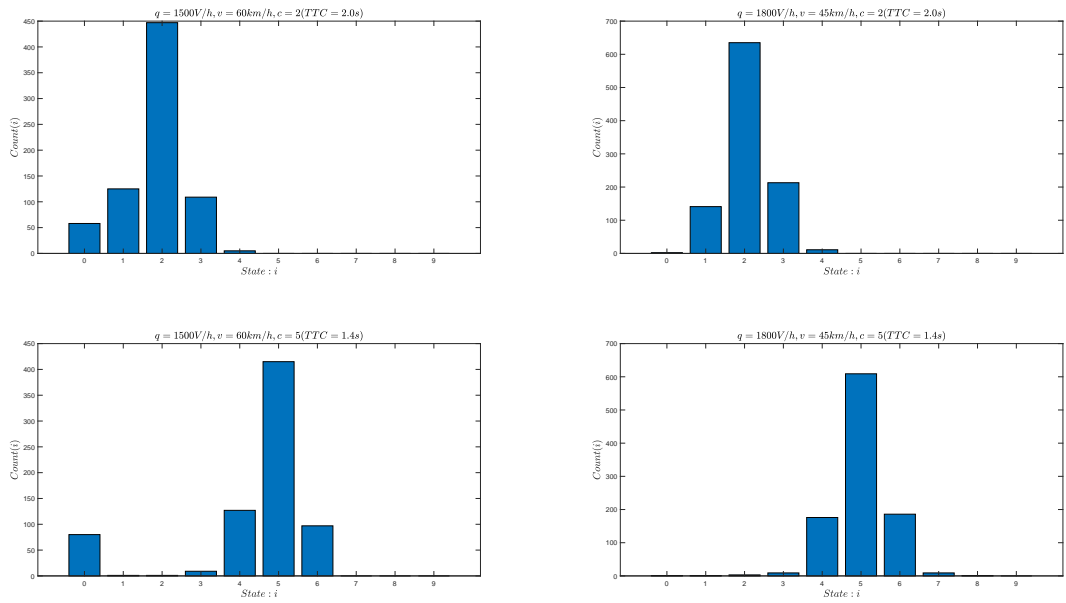}
\caption{The statistical results of the microscopic risk states of the simulation experiment.}
\label{fig:7}       
\end{figure*}

\begin{figure*}[h]
\centering
  \includegraphics[width=\textwidth]{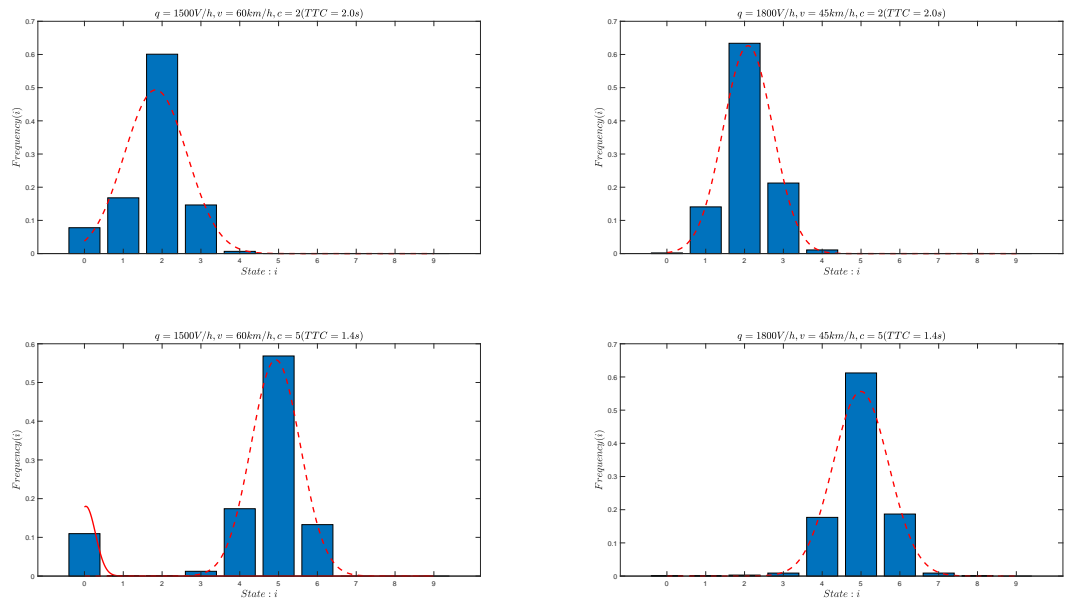}
\caption{The frequency distribution of the microscopic risk states.}
\label{fig:8}       
\end{figure*}

In this section, testing of the proposed model is carried out using autonomous driving simulation traffic data. In this simulation experiment, 4 car-following task were designed to execute on a road section between two signalized intersections, with 807 meters long and top speed limit $60 km/h$. An autonomous driving system with AEB function based on NXP i.MX8, finished 4 trip on this testing section, under different traffic flow volumes and TTC safety thresholds. Relevant data collection was obtained in the background of the simulation system and the automatic driving system.
\paragraph{}
The 4 experiment tasks are arranged by the combination of low and high flow ($q$) and high and low TTC thresholds ($c$), which are (1) $q=1500 V⁄h,c=2.0s$, (2) $q=1800 V⁄h,c=2.0s$, (3) $q=1500 V⁄h,c=1.4s$, and (4) $q=1800 V⁄h,c=1.4s$ in order. The discretization parameter $\delta$ of autonomous driving system is $1⁄15 s$ (i.e., 15 frames in each second), and the step size $\sigma$ is $0.2s$. The $\alpha$ and $\beta$ of the autonomous driving system are 0.02 and 0.34, which means that there are a 2\% probability to mistake and plan wrong, and a 34\% probability to postpone execution to the next cycle. The microscopic risk state experiment results are shown in Fig.~\ref{fig:7}.
Figure \ref{fig:7} shows that, Task (1) with low flow and high safety threshold is the safest driving styles, on the contrary, Task (4) with high flow and low safety threshold is the most dangerous driving styles (staying in state $8$ once, very close to rear-end collision state $9$). The risk state counts in Fig.~\ref{fig:7} is normalized to the frequency distribution as Fig.~\ref{fig:8}.Where, the added red line is the normal distribution trend line. 
\paragraph{}
In Fig.~\ref{fig:8}, it can be seen intuitively that, the rear-end causal effect of safety thresholds is significantly greater than the traffic flows. This view is consistent with the conclusion that human factors (including self-driving AI) are the main cause of accidents.
\section{Conclusions}
In self-driving era, automotive engineers and actuaries, just like statistical physicists, can observe and study the macro-and micro-state risks of autonomous vehicles with new generation sensing and data collection technology. The risk modeling method proposed in this paper is derived from the research results of vehicle dynamics and traffic engineering. Based on the information theory, the criteria for the optimization model are established. This new approach will significantly contribute for actuarial modeling in autonomous driving scenarios. This research has achieved, (a) {\it To measure and estimate the crash risk from a microscopic state}. A unified risk index, risk entropy, is proposed, and Discrete Markov Chain is used to describe the dynamic mechanism of the stochastic process of accidents; (b) A good causal interpretation capability. The new model can demonstrate how mistakes of autonomous driving system and road traffic conditions affect the occurrence of accidents, which is very suitable for generating simulation data to price emerging risks; and (c) Can be extended to model human driver risk. Adding aggressive driving decision-making tactics can model human driving risks \cite{yu2020a}.
\paragraph{}
In addition to this research, this new generation risk models will not only serve insurance industry, but also become a safety algorithm software of autonomous vehicles, because safety-first criteria applies to both automotive industry and insurance industry.

\begin{acknowledgements}
This study was supported by the self-driving simulation datasets of Chainlink Co. Ltd. The author thanks Dr. Long Tai Chen for his advice and support.
\end{acknowledgements}


\bibliographystyle{amsplain}  
\bibliography{bibrefs}   			% name your BibTeX data base

%\begin{thebibliography}{}
% etc
%\end{thebibliography}

\end{document}